# Non-destructive imaging of an individual protein

J.-N. Longchamp*, T. Latychevskaia, C. Escher, H.-W. Fink

Physics Institute

University of Zurich

Winterthurerstrasse 190

8057 Zurich

Switzerland

*Corresponding author:

E-mail: longchamp@physik.uzh.ch

Major category: Physics

Minor category: Biophysics and Computational Biology

**Abstract**

The mode of action of proteins is to a large extent given by their ability to adopt different conformations. This is why imaging single biomolecules at atomic resolution is one of the ultimate goals of biophysics and structural biology. The existing protein database has emerged from X-ray crystallography, NMR or cryo-TEM investigations. However, these tools all require averaging over a large number of proteins and thus over different conformations. This of course results in the loss of structural information. Likewise it has been shown that even the emergent X-FEL technique will not get away without averaging over a large quantity of molecules(1).

Here we report the first recordings of a protein at sub-nanometer resolution obtained from one individual ferritin by means of low-energy electron holography. One single protein could be imaged for an extended period of time without any sign of radiation damage(2). Since ferritin exhibits an iron core, the holographic reconstructions could also be cross-validated against TEM images of the very same molecule by imaging the iron cluster inside the molecule while the protein shell is decomposed.

**Introduction**

All conventional methods for investigating the structure of biomolecules such as proteins suffer from significant drawbacks. X-ray crystallography, NMR and cryo-electron microscopy require averaging over a large number of molecules, usually more than $10^6$ similar entities(6) and as a consequence conformational details remain undiscovered. Furthermore, these methods can only be applied to a small subset of biological molecules that either readily crystallize to be used for X-ray studies or are readily soluble and small enough (< 35 kDa) for NMR investigations.

Emerging X-ray coherent diffraction imaging (CDI) projects envision imaging the molecule instantaneously before it is destroyed by radiation damage(7). Recently, Seibert et al. succeeded in imaging a single highly symmetric 750 nm diameter mimivirus with 32 nm resolution(8). While there was the initial believe that this technique would provide high resolution images of a single biomolecule, it is now evident that averaging over a large number of molecules cannot be omitted even with the brightest source of the X-FEL projects(1, 9).

We have recently demonstrated that electrons with kinetic energies in the range of 50-200 eV do not cause radiation damage to biomolecules, enabling the investigation of an individual molecule for an extended period of time(2). Since the electron wavelength associated with this kinetic energy range is between 0.86 Å (for 200 eV) and 1.7 Å (for 50 eV), low-energy electrons have the potential for non-destructive imaging of single biomolecules and in particular individual proteins at atomic resolution.

Our low-energy electron holography experimental scheme transcribes the original idea of holography invented by Dennis Gabor(10). A sharp tungsten tip acts as field emitter of a divergent coherent electron beam(11). As illustrated in Figure 1a, the sample is brought in the path of the coherent electron wave. The electrons scattered off the object interfere with the unscattered so-called reference wave and at a distant detector the interference pattern is recorded(12). A hologram, in contrast to a diffraction pattern, contains the phase information of the object wave and the object can be unambiguously reconstructed(13).

The protein of interest needs to be free-standing in space when exposed to the low-energy electron beam. Therefore the protein is attached to a carbon nanotube suspended over a hole. Strictly speaking, arrays of small holes of 240 nm in diameter are milled in a carbon coated silicon nitride membrane by means of a focused gallium ion beam. The nanotube-ferritin complex is kept in aqueous solution of which a droplet is applied onto the holey membrane. Once the droplet has dried out some of the nanotubes remain across holes providing free-standing ferritin molecules. These molecules can then be examined in our low-energy electron microscope(12).

Prior to the described preparation procedure the carbon nanotubes undergo acid treatment in order to form carboxyl groups on the outer wall and hence disperse efficiently in ultra highly purified water. Adding a buffered solution of proteins, the latter eventually bind to the nanotubes by dipole forces, as schematically illustrated in the inset of Figure 1. This preparation does not rely on specific features of ferritin and hence is applicable to a large class of biomolecules.

Here we report the first images of ferritin at sub-nanometer resolution obtained from one individual protein by means of low-energy electron holography. The proteins could be imaged for an extended period of time without any sign of radiation damage. Ferritin is a globular protein with a molecular weight of 450 kDa and is composed of 24 subunits, the outer diameter of the protein amounts to 11-12 nm and the inner cavity exhibits a diameter of 8 nm(3-4). Up to 4500 iron atoms can be stored in the cavity and the main function of ferritin in the animal metabolism is the regulation of the iron level in the body(4). Recently, it has been shown that ferritin plays an essential role in the progression of neuronal degenerative diseases like Alzheimer or Parkinson(5). Understanding the function of ferritin in more detail by identifying its various conformations, would be an important step towards a better comprehension of the mechanism behind these diseases.

**Results**

In Figure 2a, an example of a low-energy electron hologram of ferritin molecules attached to nanotubes is presented. There are several ferritin molecules chemisorbed onto the nanotubes as evident from the hologram reconstruction. Individual ferritin molecules can be resolved, as shown in Figure 2b-c. The ability to image several molecules exhibiting different orientations and conformations in just one single-shot is a promising feature inherent to our method. The individual ferritin molecules have been monitored for more than 20 minutes without observing any change in the hologram. The corresponding total dose amounts to more than $10^9$ electrons/nm$^2$ and is at least six orders of magnitude higher than the permissible dose in TEM or X-ray examinations of biomolecules. The globular form and the size are in agreement with the structure proposed by X-ray crystallography and cryo-TEM investigations.

Due to the large amount of iron nuclei stored in the center of the molecule, ferritin can also be detected by high-energy electron imaging techniques(14-17) without any additional heavy metal staining. Though these techniques lead to radiation damage of the protein shell they allow for subsequent cross-validation in terms of presence and location of the molecule.

In Figure 3a, a hologram of a ferritin sample recorded with electrons of 57 eV kinetic energy is depicted. The presence of a bundle of carbon nanotubes as well as the globular structures on its outer wall can already be identified before the numerical reconstruction has been performed. The result of the latter is presented in Figure 3b. The right side of the nanotube corresponds to the correct representation, free of twin-image artifacts(18), as obtained by single-sideband holography reconstruction(19). A close-up of the reconstruction of an individual ferritin molecule is displayed in Figure 3c.

For cross-validation, the sample was subsequently examined in a conventional 80 kV TEM. Figure 3c displays the TEM image of the very same sample as shown in Figure 3a-c. The 8 nm iron core of the ferritin is apparent at the very same position as in the previous holographic reconstruction. While the position of the ferritin observed in the low-energy point source microscope perfectly corresponds to that observed in the TEM, the way the protein is attached to the nanotube appears to be different. In contrast to the holographic reconstruction where the protein appears as a sphere attached to an elongated structure, the TEM image displays the protein as if it was "melded" with the nanotube. The overall size of the protein observed in the TEM amounts to 9-10 nm. We associate theses differences to the radiation damage which has occurred during TEM imaging. Following the high-energy electron investigations, the sample was transferred back to our low-energy electron point source microscope; the corresponding hologram and its reconstruction are displayed in Figure 3e-f respectively. Here, the damage provoked by the 80 keV kinetic energy electrons is evident. The way the protein is now attached to the nanotubes coincides with the TEM results.

**Discussion**

We have reported the very first non-destructive investigation of an individual protein by means of low-energy electron holography. The successful imaging of a single protein is cross-validated by TEM examination. The sample preparation method can be applied to a broad class of molecules.

While the spatial resolution in the holographic reconstructions presented here is already below the nanometer scale, it is now an experimental challenge to reach a resolution close to the employed wavelength with an advanced low-energy electron holographic setup.

Besides this engineering challenge of improving interference resolution in holography, another route towards higher resolution is to combine holography with coherent diffraction imaging using low-energy electrons(20). Essential ingredients like a micron sized electron lens(21) and the ability to record coherent diffraction patterns(22) are already at hand for the pursuit of single molecule structural biology at atomic resolution.

## Methods

**Sample preparation.** COOH-functionalized double-walled carbon nanotubes from the company NanoLab were dispersed to a concentration of 0.2 mg/ml by sonication for 15 min in ultra highly purified water. Subsequently, ferritin proteins in a MES buffer (pH 6.5) were added to the carbon nanotubes solution, with a final concentration of 0.5 nM. The ferritin-nanotube solution was kept under gentle stirring overnight.

SiN membranes (50 nm thickness) from the company Silson were first coated with a 5 nm thick carbon layer. Subsequently, holes of 200 nm in diameter were milled in the membranes by means of a focused Ga ion beam. Thereafter, the membranes were coated again with 20 nm carbon layers on both sides.

Next, the membranes were hydrophylized by exposition to ozone in an UV-ozone oven for 20 min. Before 1 $\mu$l of the ferritin-nanotubes solution was applied and air-dryed on the membrane.

**Holography.** The detector system of our low-energy electron point source microscope consists of a multi-channel plate (MCP) and a phosphor screen directly mounted on a fiber optic plate (FOP). Both, MCP and Phosphor screen-FOP are 75 mm in diameter. A 8000x6000 pixels CCD camera was used to record the holograms. The overall resolution of the detector system is around 20 $\mu$m.

An electro chemically etched single-crystal <111>-tungsten tip was used as point source of coherent low-energy electrons. The holograms were recorded with a total electron current of 50 nA.

**Acknowledgements**

We would like to thank Jennifer Clark for her help with sample preparation and the Swiss National Science Foundation for its financial support.

**Author Contributions**

J.N.L. conducted the holographic and TEM investigations. T.L. performed the hologram reconstructions. J.N.L. and C.E. worked on the sample preparation and H.W.F. contributed to the design of the experimental setup. All authors discussed the results and implications and jointly wrote the manuscript.

**Figure legends**

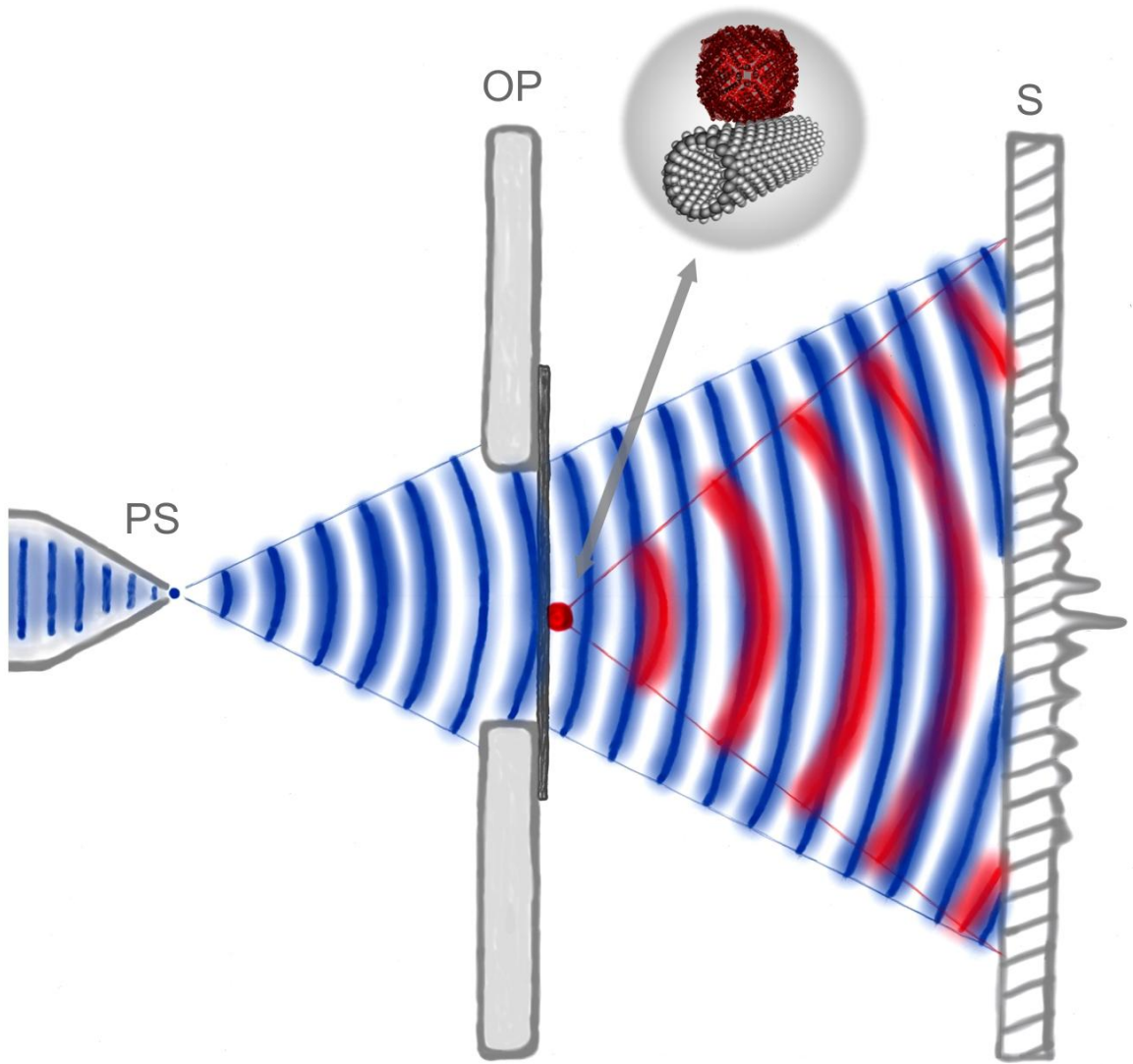

**Figure 1: Scheme for recording the low-energy electron hologram of a protein.** Conduction electrons confined in a pointed W(111) single crystal wire are field emitted into vacuum at an atomic-sized emission area providing a coherent low-energy electron point source (PS). At the less than 1 micron distant object-plane (OP), part of the coherent electron wave is scattered by a ferritin attached to a carbon nanotube constituting the object wave indicated in red. At a distant detector screen (S), the far-field interference pattern between object- (red) and reference-wave (blue) - the hologram - is recorded and its digital record is subject to the numerical reconstruction of the protein.

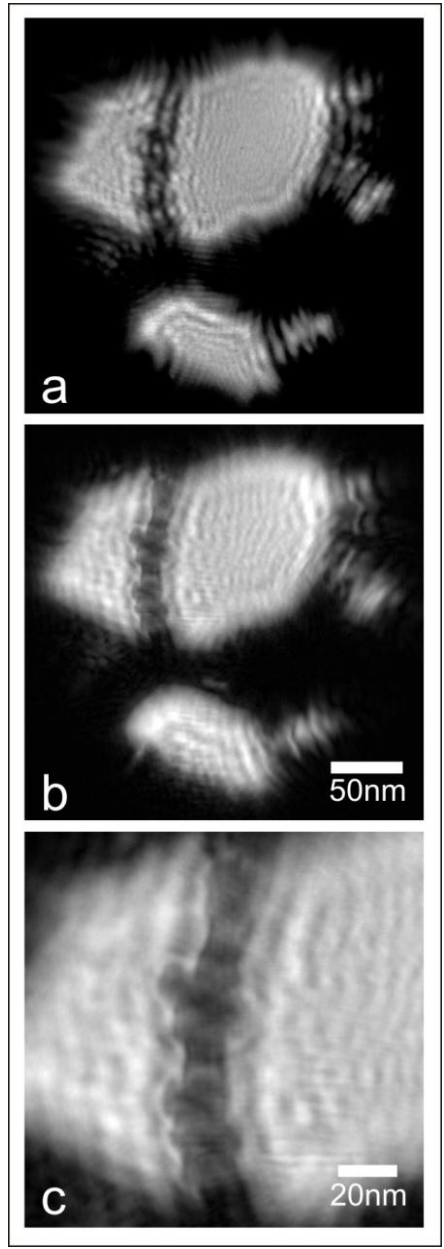

**Figure 2: Low-energy electron hologram of ferritin and its numerical reconstruction.** **a** 53eV kinetic energy electron hologram of several ferritin molecules attached to a bundle of carbon nanotubes. **b** reconstruction of **a**. **c** close-up of **b.**

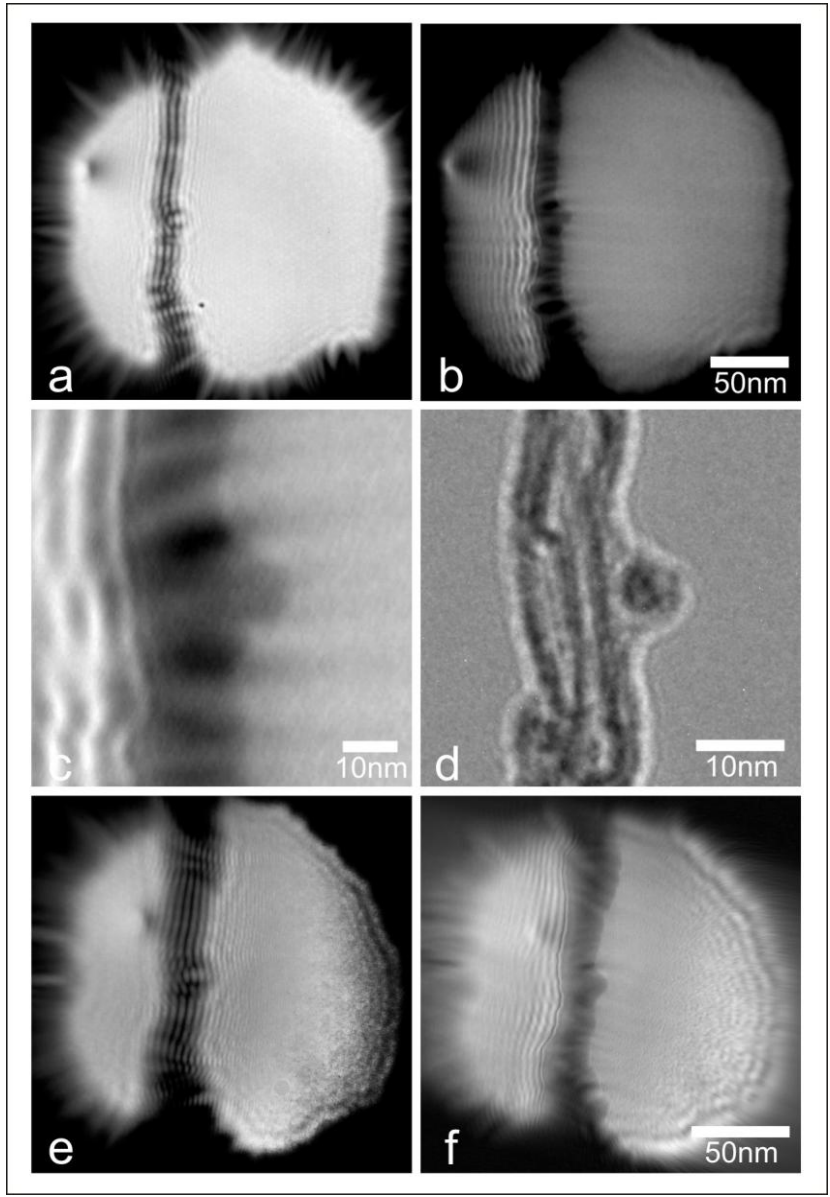

**Figure 3: Control experiments using a TEM. a** 57eV kinetic energy electron hologram of individual ferritin molecules attached to a bundle of carbon nanotubes. **b** side-band holography reconstruction of **a**. **c** close-up of **b**. **d** TEM image of the very same ferritin molecule. **e** 45eV kinetic energy electron hologram of the very same sample as in **a** but after the TEM investigations. **f** side-band holography reconstruction of **e**.